\documentstyle[twoside, epsf]{article}
\input{epsf.sty}
\input ibvs2.sty

\begin{document}

\IBVShead{5816}{04 February 2008}

\IBVStitle{Does the period of BE Lyncis really vary?}

\IBVSauth{SZAK\'ATS, R., SZAB\'O, Gy.M., SZATM\'ARY, K.}

\IBVSinst{Dept. Experimental Physics \& Astronomical Obs., Univ. of Szeged,
6720 Szeged D\'om t\'er 
9, Hungary \newline e-mail: szgy@titan.physx.u-szeged.hu}

\SIMBADobjAlias{BE Lyn}{}
\IBVStyp{HADS}
\IBVSkey{analysis}
\IBVSabs{New photometric series from 6 epochs of BE Lyncis are presented.}
\IBVSabs{With template curve fitting we re-determined the $O-C$ for BE Lyncis.}
\IBVSabs{The phase shift diagram is apparently constant, 
disproving the suspected period variations of BE Lyn.}

\begintext

The first period variation analyses { (Liu et al., 1991, Tang et al., 1992, Liu \&{} Jiang., 1994)}
of the HADS star BE Lyncis (m$_V\approx 8.8$ mag, $P=$0\fday 0958695)
indicated a parabolic fit of the O$-$C. Kiss \& Szatm\'ary (1995)
suggested { the presence of period variations possibly due to a} companion. 
Derekas et al. (2003, D03) re-analyzed
the available data, and disproved the light-time { hypothesis}. They also
noted that the scatter of the points in the $O-C$ diagram was slightly higher 
than the accuracy of
individual data points, which might refer to microvariability. { However, both} 
Rodr\'{\i}guez et al. (1996) and D03 failed to detect additional { frequency components}. 
Later, Fu \& Yiang (2005) revived the binary hypothesis again.
The purpose of this paper is to { test} whether there is cyclic phase 
modulation in the light curve of BE Lyn that may refer to a light-time effect. We 
present new times of maxima from 6 { epochs} between 2003--2006, and { re-analyze}
the available light curve data { with} { phase shift analysis (e.g. Jurcsik et al., 2001)} 
using template curve fitting.

\begin{table}[h]
\begin{center}
\begin{tabular}{cccccc}
Date   & HJD &      length & number     &  instrument     &         filter\\
       & (first point) & (hour) &  of points \\
\hline
2003.12.08 & 2452981.51    &  4.03    & 174 &        1.0 RCC&   V\\
2004.04.22 & 2453119.47   &   2.98    & 281 &        1.0 RCC&   V\\
2006.10.18 & 2454027.47   &   4.97    & 567 &        0.4 N&    V\\
2006.10.19 & 2454028.42   &   5.90    & 668 &        0.4 N&    V\\
2006.10.27 & 2454036.44   &   5.74    & 568 &        0.4 N&    V\\
2006.10.30 & 2454039.42   &   6.47    & 833 &        0.4 N&    V\\
\hline       
\end{tabular}
\caption{The log of new observations}
\end{center}
\end{table}

\IBVSfig{10cm}{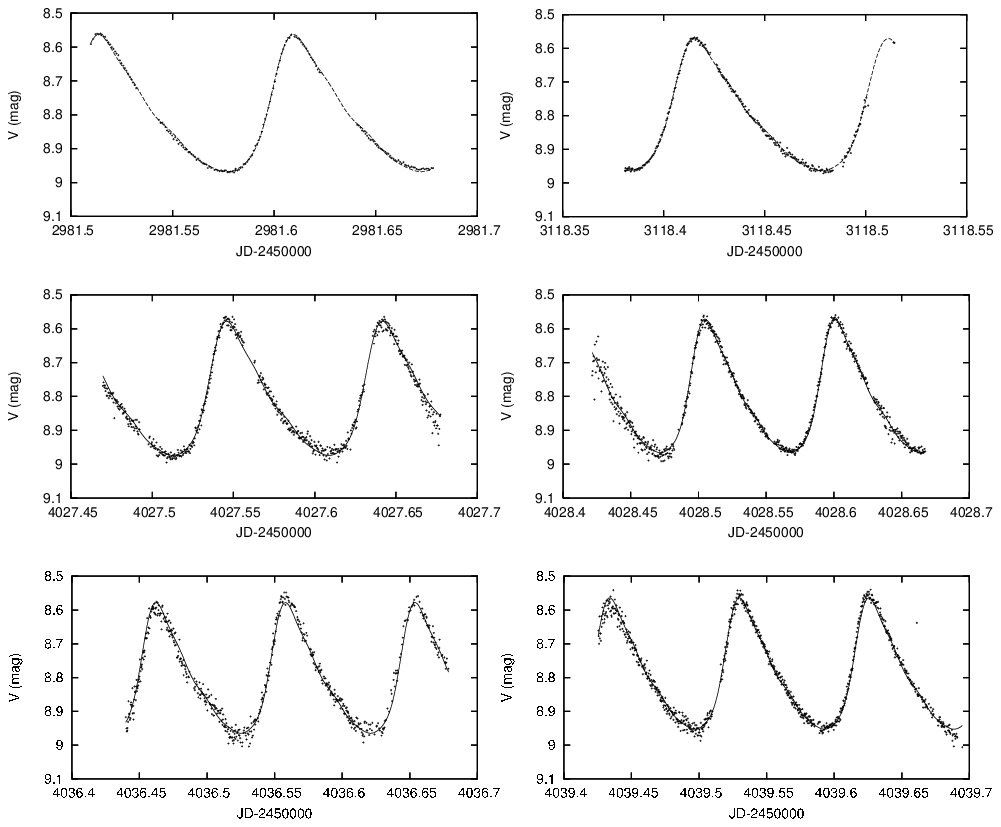}{New light curves of BE Lyncis}

We took new CCD observations of BE Lyn from two different sites. First, we used the the 1-meter 
RCC telescope of Konkoly Observatory, located at the Piszk\'estet\H{o} Mountain 
Station (1.0 RCC). The typical integration time was 8
s for Johnson V filter. In October, 2006 further observations
were made with the 40 cm Newton telescope (0.4 N) of the University of Szeged, Dept. of Experimental
Physics and Astronomical Observatory. Instrumental magnitudes were taken with 
Johnson V filter and with typically 10 second exposures.
The log of observations is { listed} in Table 1, the light curves are shown in
Fig. 1. The light curves are published electronically on the IBVS site
(http://www.konkoly.hu/cgi-bin/IBVSdatatable?5816-t2).

In addition to the new observations, we collected light curves obtained between 1987  and 2007 from the literature.
A template curve was determined by a 5-th order Fourier polynomial fit to all observations,
$$f(t)=a_0+A\sum_{k=1}^5 \left( a_k \sin k\phi + b_k \cos k\phi \right),$$
where { $A$ is the relative amplitude, $a_0$ is the mean brightness and $\phi=2 \pi t/P$.}
The resulted coefficients defining the template curve are:
$a_0=8.8128$,
$a_1=0.0740$,
$b_1=-0.1578$,  
$a_2=0.0523$,   
$b_2=0.0151$,
$a_3=0$,
$b_3=0.0207$,
$a_4=-0.0097$,
$b_4=0$,
$a_5=-0.0034$,
$b_5=-0.0041$. 
{ If $A=1$, this template curve has a total amplitude of $0.395$ mag.}
The template curve was { then} fitted to the individual observing runs allowing a { slight} global 
phase shift. { Because the observed light curve varied slightly, the $A$ amplitude parameter and the $a_0$ mean
brightness was also fitted as a free parameter.}
The time of maximum of the { best-fit} { model light} curve can be similarly { evaluated}
as the $O-C$, { using} calculated moments of maxima as $C=2449749.4651+0.09586952 \cdot E $. 
We determined a refined period as $P=0\fday 09586952\pm 0\fday 00000003$ at 3--$\sigma$ confidence level.

{ Photometric data were available for us from Oja, 1987; Rodr\'\i{}guez et al., 1990, Kiss \&{} Szatm\'ary, 1995; D03 and the measurements published here.}
We show the $O-C$ diagram of maxima for all published data in the upper panel of Fig. 2 with open circles.
The lower panel shows the phase shift diagram ($O-C$ of the fitted template curves) 
{ from the available photometries suitable for re-analysis} (for comparison, these points are
highlighted with filled circles in the upper panel, too). The errors were calculated from the correlation matrices of the parameters. 

All {new and} re-determined times of maxima and amplitudes are available at the IBVS site (http://www.konkoly.hu/cgi-bin/IBVSdatatable?5816-t3). 
This table also includes the moments of maxima from the archive time series even if the data were not available for the present analysis, in this case the appropriate columns are vacant.

\IBVSfig{11cm}{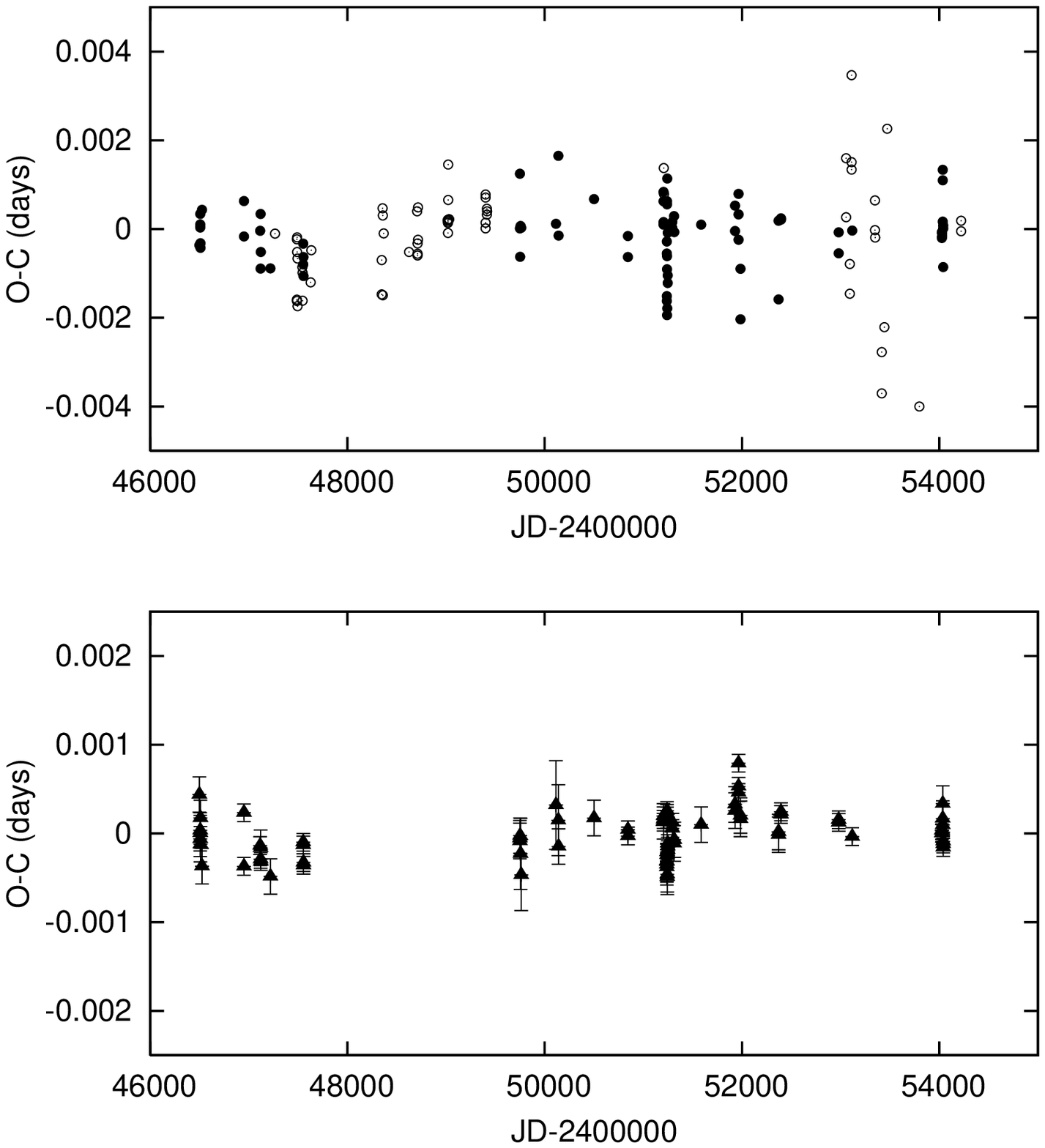}{Upper panel: the $O-C$ of BE Lyn from times of maxima. Open circles show all published data,
filled circles show photometries involved into phase shift analysis. Lower panel: The phase shift diagram of BE Lyncis. Note that the
$O-C$ axis has half scale as in the upper panel.}

The re-determined times of maxima show only little variation,
all  data points are practically $0$ within some 10 second accuracy.
This strongly suggests that there is no variation in the global phase of the
light curves. On the other hand, we confirm that the scatter of the classical $O-C$
diagram is too high to be a single artefact (as also noted by D03). 
Thus, we suggest that the phase of the maximum brightness 
varies slightly, leading to the observed behaviour of the $O-C$ of light maxima.

Amplitude variations are present in the data set
with a range of about {0.03 mag} (Fig. 3), as first noted by Rodr\'\i{}guez et al. (1996). 
We revisited the nature of the amplitude variation using
Fourier-analysis, and we confirm that it is not periodic.
The majority of the observed amplitudes is between 0.375 and 0.415 mag.
This may be caused either by the different instrumental systems or simply by
the extinction corrections, which lacks in some cases.
\IBVSfig{5.7cm}{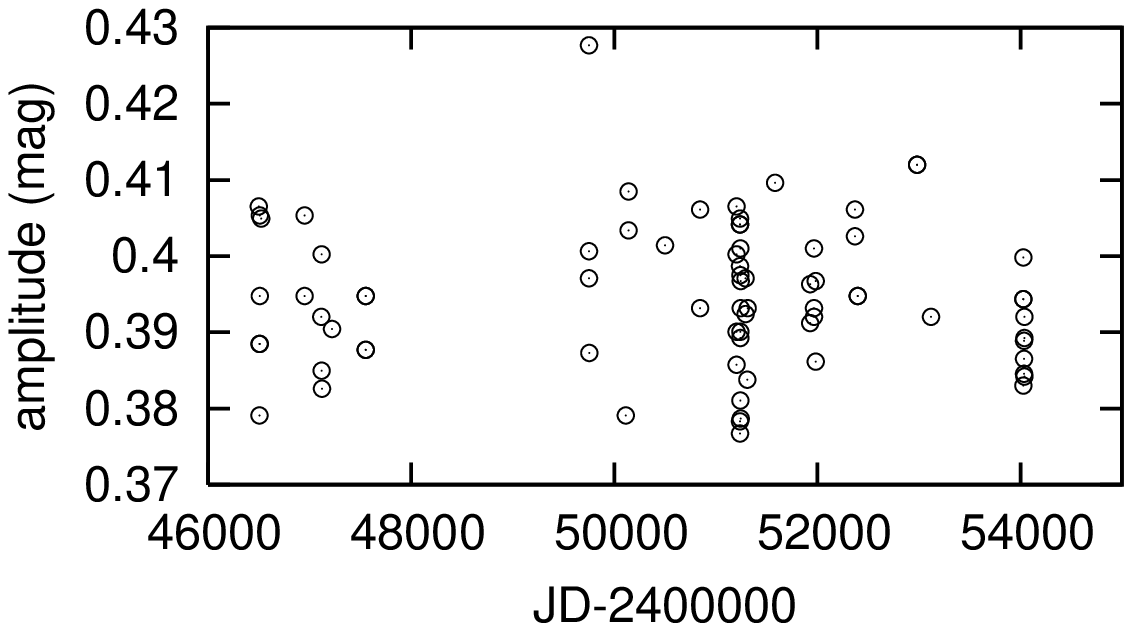}{The variation of the total amplitude of BE Lyncis from template fitting.}
{ The correlated variation of the light curve shape and the amplitude is a  known property
of the Blazhko RR Lyrae stars (Jurcsik et al. 2005).
The suspected variation of the amplitude and the light curve shape of BE Lyncis might suggest
that they also vary in a correlated way. To test this in BE Lyn, we plotted the phase of the maximum 
vs. the amplitude, and we found them to be uncorrelated.}

{\bf Acknowledgements}
The research was supported by the Hungarian OTKA Grant T042509. GyMSz was supported by
the Bolyai J\'anos Research Fellowship of the Hungarian Academy of Sciences. 
The telescope time at the Konkoly Observatory is acknowledged.

\references

Derekas, A., Kiss, L.L., Sz\'ekely, P., Alfaro, E.J., et al., 2003, {\it A\&{A}}, { 402}, 733

Fu, J.N., Jiang, S.Y., 2005, {\it ASP Conf. Ser.}, { 335}, 293

H\"ubscher, J., 2005, {\it IBVS}, No. 5643

H\"ubscher, J., 2007, {\it IBVS}, No. 5802

H\"ubscher, J., Paschke, A., Walter, F.,  2005a, {\it IBVS}, No. 5657

H\"ubscher, J., Lange, Th., Paschke, A., Vohla, F., Walter, F.,  2005b, {\it BAV Mitteilungen}, Nr. 174

H\"ubscher, J., Paschke, A., Walter, F.,  2006, {\it IBVS}, No. 5731

Jurcsik, J., Clement, C., Geyer, E. H., Domsa, I., 2001, {\it AJ}, { 121}, 951

Jurcsik, J., S\'odor, \'A., V\'aradi, M., et al., 2005, {\it A\&{A}}, { 430}, 1049

Kiss, L.L., Szatm\'ary, K., 1995, {\it IBVS}, No. 4166

Klingenberg, G., Dworak, S.W., Robertson, C.W., 2006, {\it IBVS}, No. 5701

Liu, Y., Jiang, S., Cao, M., 1991, {\it IBVS}, No. 3607

Liu, Zh., Jiang, S., 1994, {\it IBVS}, No. 4077

Oja, T., 1987, {\it A\&{A}}, { 184}, 215

Rodr\'{\i}guez, E., L\'opez de Coca, P., Rolland, A., Garrido, R., 1990, {\it Rev. Mex. Astron. Astrofis.}, { 20}, 37

Rodr\'{\i}guez, E., L\'opez de Coca, P., Mart\'in, S., 1996, {\it A\&{A}}, { 307}, 539

Tang, Q., Yang, D., Jiang, S., 1992, {\it IBVS}, No. 3771

Wunder, E., Wieck, M., Garzarolli, M., 1992, {\it IBVS}, No. 3791

\endreferences


\end{document}